\documentstyle[epsfig]{elsart}
\newcommand{\lsim}{\raisebox{-0.55ex}{$\stackrel{\displaystyle <}{\sim}$}}

\newcommand{\diff}[2]{\frac{\partial #1}{\partial #2}}

\newcommand{\rmf}[1]{{\rm #1}}
\begin{document}
\begin{frontmatter}

%
\title{Scaling behavior of $(N_{\rm ch})^{-1}\rmf dN_{\rm ch}/\rmf d\eta$ at $\sqrt{s_{\rm NN}} = 130\ {\rm GeV}$ by PHOBOS Collaboration and its analyses in terms of stochastic approach}

\author[Shinshu]{M. Biyajima,}
\author[Shinshu]{M. Ide,}
\author[Toba]{T. Mizoguchi}
\author[Matsu]{and N. Suzuki}

\address[Shinshu]{Department of Physics, Faculty of Science, Shinshu University, Matsumoto 390-8621, Japan}
\address[Toba]{Toba National College of Maritime Technology, Toba 517-8501, Japan}
\address[Matsu]{Matsusho Gakuen Junior College, Matsumoto 390-1295, Japan}

%
\begin{abstract}
Recently interesting data on $\rmf dN_{\rm ch}/\rmf d\eta$ in Au-Au collisions ($\eta=-\ln \tan (\theta/2)$) with the centrality cuts have been reported by PHOBOS Collaboration. Their data are usually divided by the number of participants (nucleons) in collisions. Instead of this way, using the total multiplicity $N_{\rm ch} = \int (\rmf dN_{\rm ch}/\rmf d\eta)\rmf d\eta$, we find that there is scaling phenomenon among $(N_{\rm ch})^{-1}\rmf dN_{\rm ch}/d\eta = \rmf dn/d\eta$ with different centrality cuts at $\sqrt{s_{\rm NN}} = 130$ GeV. To explain this scaling behavior of $\rmf dn/\rmf d\eta$, we consider the stochastic approach called the Ornstein-Uhlenbeck process with two sources. Moreover, comparisons of $\rmf dn/\rmf d\eta$ at $\sqrt{s_{\rm NN}} = 130$ GeV with that at $\sqrt{s_{\rm NN}} = 200$ GeV have been made. A possible detection method of the quark-gluon plasma (QGP) thorough $\rmf dn/\rmf d\eta$ is presented.
\end{abstract}

\end{frontmatter}

%
\section{Introduction}
Recently PHOBOS Collaboration has published an interesting data on $\rmf dN_{\rm ch}/\rmf d\eta$ ($\eta=-\ln \tan (\theta/2)$)\footnote{
\begin{eqnarray*}
  y &=& \frac 12 \ln \frac{E+p_z}{E-p_z} = \tanh^{-1} \left(\frac{p_z}E\right) \approx -\ln\tan(\theta/2) \equiv \eta\, .\\
  \eta &=& \frac 12 \ln \frac{p+p_z}{p-p_z}\, ,\ {\rm and}\ \frac{\rmf dn}{\rmf d\eta} = \frac pE \frac{\rmf dn}{\rmf dy}\, .
\end{eqnarray*}
} 
in Au-Au collisions at $\sqrt{s_{\rm NN}} = 130$ GeV \cite{Back:2001bq}. The authors of ref. \cite{Back:2001bq} have calculated the following quantity,
\begin{eqnarray}
\frac 1{\langle N_{\rm part}\rangle/2}\frac{\rmf dN_{\rm ch}}{\rmf d\eta} = f(N_{\rm part},\,N_{\rm coll},\,\eta)\, ,
\label{eq1}
\end{eqnarray}
where $N_{\rm part}$ and $N_{\rm coll}$ mean the number of participants (nucleons) and number of collision particles in Au-Au collisions. It depends on the centrality cuts. The function $f(N_{\rm part},\,N_{\rm coll},\,\eta=0)$ is an increasing function, as $N_{\rm part}$ increases.

\begin{figure}[htb]
  \centering
  \epsfig{file=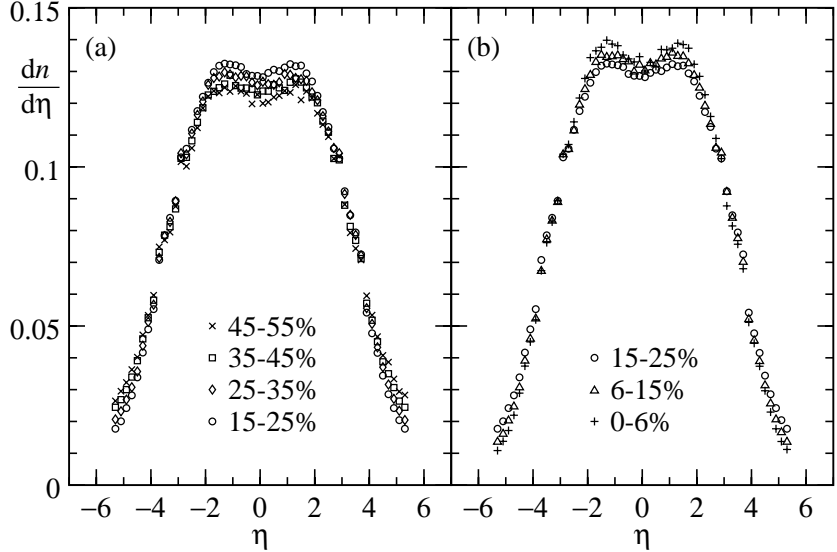,scale=0.6}
  \caption{Two sets of $\rmf dn/\rmf d\eta$ with different centrality cuts \cite{Back:2001bq}. Data with 15-25\% are used in both Figures, for the sake of comparisons.}
\label{fig1}
\end{figure}

In this paper, instead of eq. (\ref{eq1}), we consider the following physical quantity,
\begin{eqnarray}
  \frac 1{N_{\rm ch}}\frac{\rmf dN_{\rm ch}}{\rmf d\eta} = \frac{\rmf dn}{\rmf d\eta}\, ,
\label{eq2}
\end{eqnarray}
where $N_{\rm ch} = \int (\rmf dN_{\rm ch}/\rmf d\eta)\rmf d\eta$, and $\int (\rmf dn/\rmf d\eta)\rmf d\eta = 1$. In Fig. \ref{fig1}, two sets of $\rmf dn/\rmf d\eta$ are shown. They suggest us that there is scaling among $\rmf dn/\rmf d\eta$'s with different centrality cuts. Thus $\rmf dn/\rmf d\eta$ is named a kind of the probability density, because $\rmf dn/\rmf d\eta = f(\eta)$, where $\rmf dn/\rmf d\eta$ is function of $\eta$ only   This fact probably implies that the stochastic approach is available in analyses of $\rmf dn/\rmf d\eta$\footnote{
Multiplicity distributions in high energy collisions, i.e., $P(n,\, \langle n\rangle)$'s are the probability distributions, which are function of $n$ and $\langle n\rangle$. It is known that the KNO scaling functions,
$$
\lim_{n,\,\langle n\rangle \to \infty} \langle n\rangle P(n) = \psi (z=n/\langle n\rangle)
$$
are described by solutions of various Fokker-Planck equations \cite{Koba:1972ng,Biyajima:1984qu,Biyajima:1984qv}. 
}
\footnote{
Dokshitzer has calculated the generalized gamma distribution in QCD \cite{Dokshitzer:1993dc}
$$
P(z=n/\langle n\rangle) \approx \frac{2\mu^2}z \frac{(Dz)^{3\mu/2}}{\sqrt{2\mu\gamma}}\exp \left[-(Dz)^{\mu}\right]\: ,
$$
where $z$, $\mu$, $D$, $\gamma$ are KNO scaling variable, $1-\gamma = 1/\mu$, a parameter, the anomalous dimension in QCD, respectively. This is a steady solution of the following Fokker-Planck Equation
$$
\diff Pt = -\diff{}z \left[\left(d+\frac 12Q\right)z + bz^{1+\gamma}\right]P + \frac 12Q^2\diff{^2}{z^2}[z^2P]\: .
$$
}

Contents of the present paper are as follows. In the second paragraph, the Fokker-Planck equation is considered. In the third one, concrete analyses are presented. In the final one, concluding remarks are given.

%
\section{Stochastic approach to $\rmf dn/\rmf d\eta$}
It is well known that the rapidity ($y\approx \eta$) is a kind of the velocity. Moreover, in collisions leading particles in the beam and target nuclei, i.e., nucleons in gold at RHIC experiments, collide each other, lose their energies and emit various particles. In other words, there are fluctuations in the rapidity space or the pseudo-rapidity ($\eta$). Thus we would like to assume that a damping law governs the rapidity space
\begin{eqnarray}
  \frac{\rmf dX}{\rmf dt} = - \gamma X + f_{\rm W}(t)\, ,
\label{eq3}
\end{eqnarray}
where $X=y$ or $\eta$, $\gamma$ and $f_{\rm W}(t)$ are the frictional coefficient and the white noise, respectively. From eq. (\ref{eq3}), we can derive the following Fokker-Planck equation,
\begin{eqnarray}
  \diff{P(x,\, t)}t = \gamma \left[\diff{}x + \frac 12\frac{\sigma^2}{\gamma}\diff{^2}{x^2}\right] P(x,\, t)\, ,
\label{eq4}
\end{eqnarray}
where $\sigma$ and $P(x,\, t)$ are the dispersion and the probability density, respectively. Equation (\ref{eq4}) is called the Ornstein-Uhlenbeck (O-U) process \cite{Kampen:1981,Goel:1974}. To look for the solutions of eq. (\ref{eq4}), we have to assume the source functions at ``time'' $t=0$. The typical solution with $\delta (\eta - \eta_0) = P(\eta,\,t=0)$ is given as 
\begin{eqnarray}
  P(\eta|\eta_0,\, t) = \frac 1{\sqrt{2\pi V^2(t)}}\exp\left[-\frac{(\eta-\eta_0e^{-\gamma t})^2}{2V^2(t)}\right]\, ,
\label{eq5}
\end{eqnarray}
where $V^2(t) = (\sigma^2/2\gamma)(1-e^{-2\gamma t})$ and $\eta_0$ is the initial rapidity.

\begin{figure}[htb]
  \centering
  \epsfig{file=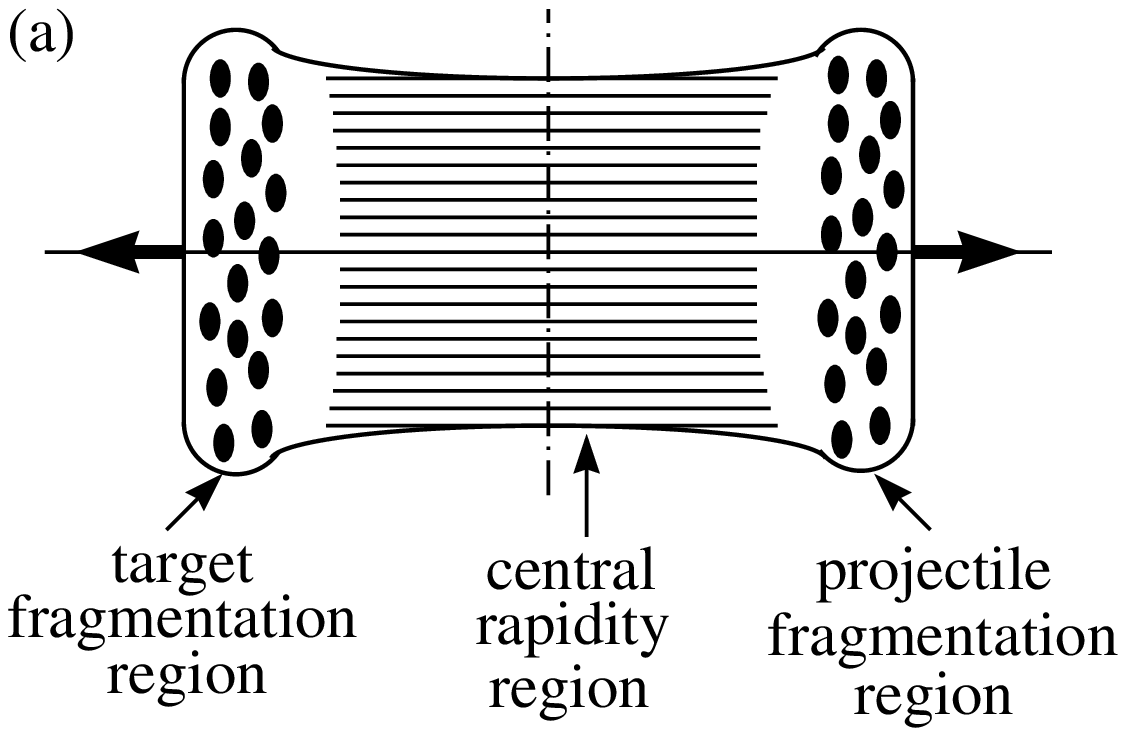,scale=0.6}
  \epsfig{file=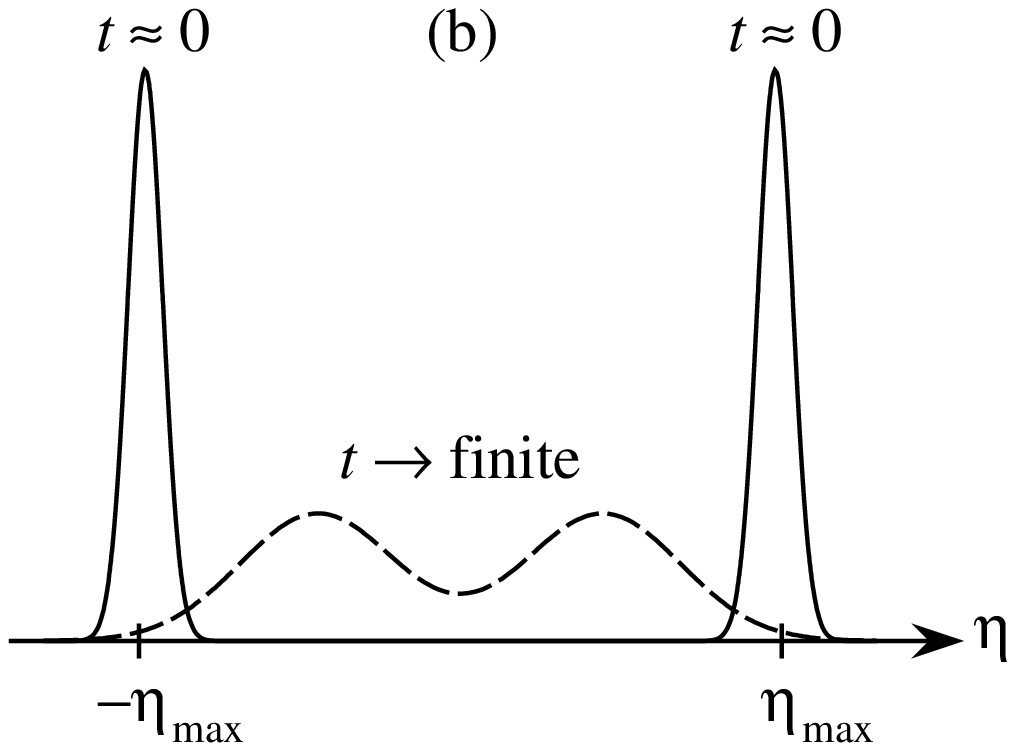,scale=0.6}
  \caption{(a) Simplified picture for A-A collisions. The black circles mean nucleons. (b) Evolution of eq. (\ref{eq6}) with two sources at $\eta_{\rm max}$ and $-\eta_{\rm max}$.}
\label{fig2}
\end{figure}

In Fig. \ref{fig2}(a), we depicted a simplified picture of heavy-ion collisions. According to Fig. \ref{fig2}(a), a model of two sources $0.5\times \delta (\eta - \eta_{\rm max})$ and $0.5\times \delta (\eta + \eta_{\rm max})$ at $t=0$ seems to be reasonable. As our present model is very simple, $0.5\times N_{\rm ch}$ particles have been produced at $\eta_{\rm max}$ and the same $0.5\times N_{\rm ch}$ particles at $-\eta_{\rm max}$ at $t=0$. In this case we have the following expression which is the sum of two solutions,
\begin{eqnarray}
  P(\eta|\eta_{\rm max},\, t) &=& 
\frac 1{\sqrt{8\pi V^2(t)}}\left\{
\exp\left[-\frac{(\eta+\eta_{\rm max}e^{-\gamma t})^2}{2V^2(t)}\right]\right . 
\nonumber\\
 &&\qquad\qquad\quad\left .+ \exp\left[-\frac{(\eta-\eta_{\rm max}e^{-\gamma t})^2}{2V^2(t)}\right]\, \right\}\, .
\label{eq6}
\end{eqnarray}
The evolution of eq. (\ref{eq6}) is shown in Fig. \ref{fig2}(b).

%
\section{Analyses of $\rmf dn/\rmf d\eta$ by means of eq. (\ref{eq6})}
\begin{figure}[htb]
  \centering
  \epsfig{file=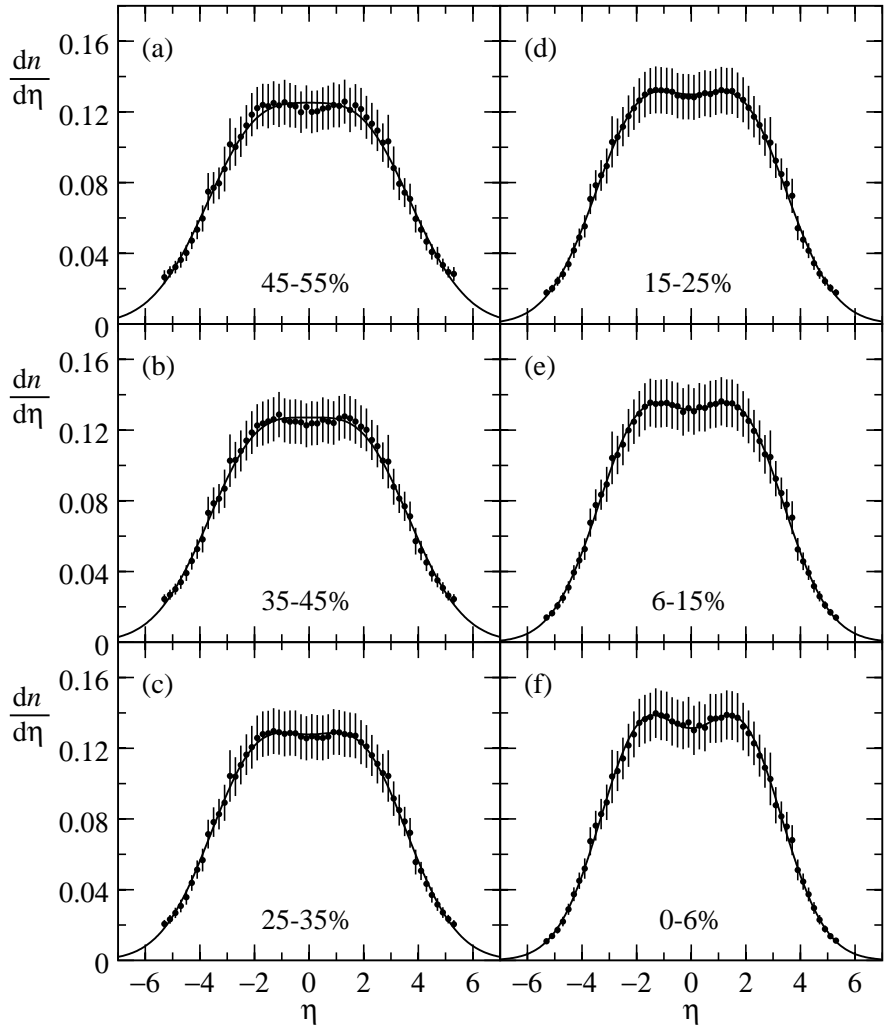,scale=0.6}
  \caption{Analyses of $\rmf dn/\rmf d\eta$ by eq. (\ref{eq6}). See Table \ref{table1}.}
\label{fig3}
\end{figure}

\begin{figure}[htb]
  \centering
  \epsfig{file=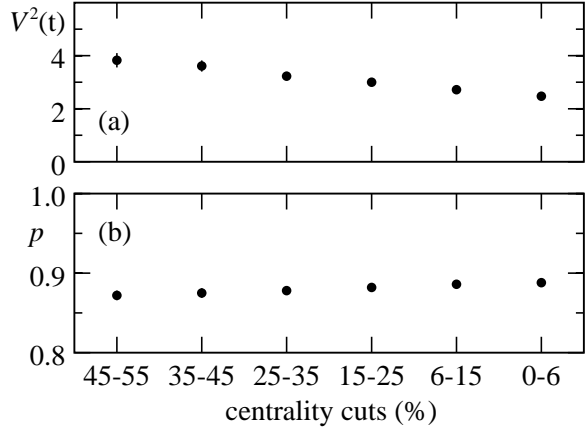,scale=0.6}
  \caption{Dispersion $V^2(t)$ and $p$ of Fig. \ref{fig3} and Table \ref{table1}.}
\label{fig4}
\end{figure}

By making use of eq. (\ref{eq6}), we can analyse $\rmf dn/\rmf d\eta$ shown in Fig. \ref{fig1}. Our results are shown in Fig. \ref{fig3} and Table \ref{table1}. In Fig. \ref{fig4}, we examine whether or not the dispersion $V^2(t)$ and $p=1-e^{-2\gamma t}$ depend on the centrality cuts. As is seen in Figs. 3 and 4, the scaling behavior among $\rmf dn/\rmf d\eta$'s at $\sqrt{s_{\rm NN}} = 130$ GeV is explained by eq. (\ref{eq6}) with small changes in the dispersion $V^2(t)$. It can be said that the scaling behavior  is explained by the O-U process with two sources at the beam ($y_{\rm B}$ or $\eta_{\rm max}$) and target ($y_{\rm T}$ or $-\eta_{\rm max}$) rapidities. Of course, it is obvious that the single source cannot explain it.
%
%
\begin{table}[htb]
\begin{center}
\caption{Estimated parameters in our analyses by eq. (\ref{eq6}) with two sources. Evolution of eq. (\ref{eq6}) is stopped at minimum $\chi^2$'s. $\delta p = 0.006\sim 0.004$.}
\label{table1}
\begin{tabular}{ccccccc} \hline
Fig. \ref{fig3} & (a) & (b) & (c) & (d) & (e) & (f)\\ \hline
centrality (\%) & 45-55 & 35-45 & 25-35 & 15-25 & 6-15 & 0-6 \\
$p$ & 0.872$\pm \delta p$ & 0.875$\pm \delta p$ & 0.878$\pm \delta p$ & 0.882$\pm \delta p$ & 0.886$\pm \delta p$ & 0.888$\pm \delta p$\\
$V^2(t)$ & 3.83$\pm$0.27 & 3.61$\pm$0.21 & 3.23$\pm$0.16 & 3.00$\pm$0.13 & 2.72$\pm$0.10 & 2.47$\pm$0.08\\
$\langle N_{\rm part}\rangle$ & --- & 93 & 135 & 197 & 270 & 340\\
$N_{\rm ch}$ & 662$\pm$10 & 1056$\pm$16 & 1582$\pm$23 & 2270$\pm$34 & 3199$\pm$49 & 4070$\pm$63\\
$\chi^2/n.d.f.$ & 8.61/51 & 7.63/51 & 5.88/51 & 5.35/51 & 3.57/51 & 3.82/51\\ \hline
\end{tabular}
\end{center}
\end{table}

%
\section{Concluding remarks}
First of all, it can be stressed that there is the scaling among $\rmf dn/\rmf d\eta$'s with various centrality cuts at $\sqrt{s_{\rm NN}} = 130$ GeV.

Second the scaling behavior of $\rmf dn/\rmf d\eta$ is described by the solutions of the Fokker-Planck equation,i.e., eq. (\ref{eq6})\footnote{
To estimate the ``thermalization time'' of the QGP, Hwa has considered the Fokker-Planck equation for the motion of the quarks and gluons in nuclei \cite{Hwa:1985fj}. See also ref. \cite{Chakraborty:1984ha} in which the Wiener process is considered for the quarks and gluons.
}. 
This suggests that $\rmf dn/\rmf d\eta$ with the centrality cut 0-6\% do not show singular /or particular phenomenon relating to signatures of the Quark-Gluon Plasma (QGP). Of course, we should pay our attention that we are handling the averaged quantity in statistics. At present, however, it is difficult to conclude that the QGP is created, and the signature from the QGP are washed out by the strong interactions between hadrons, if the QGP is created.

Here we should carefully observe Fig. \ref{fig1}. As is seen in Fig. \ref{fig1}, there are small differences in $\rmf dn/\rmf d\eta$'s over the range $|\eta| \lsim 2$ with centrality cut 0-6\% and others. To explore the differences more carefully, we need $\rmf dn/\rmf d\eta$ with smaller centrality cuts, 0-3\% $\sim$ 0-5\%. 
\begin{figure}[htb]
  \centering
  \epsfig{file=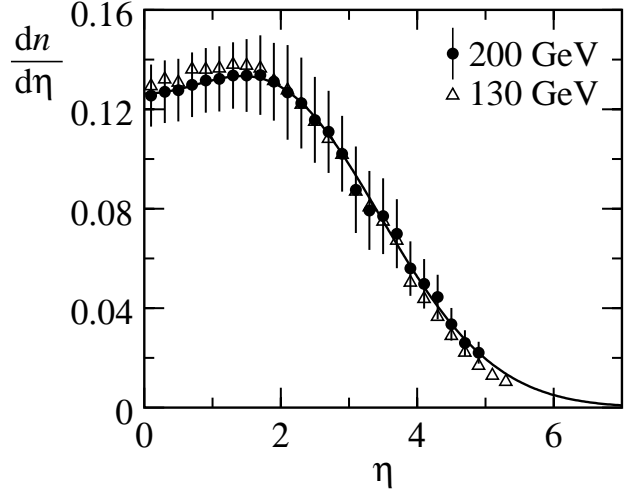,scale=0.75}
  \caption{Comparisons of $\rmf dn/\rmf d\eta$ with the centrality cut 0-6\% at $\sqrt{s_{\rm NN}} = 130$ GeV and 200 GeV. Solid line is obtained for latter energy: $p = 0.879\pm 0.007$, $V^2(t) = 2.67\pm 0.24$, $\chi^2/n.d.f = 0.63/22$.}
\label{fig5}
\end{figure}

Moreover, we can add the following fact. Very recently, PHOBOS Collaboration has reported the data on $\rmf dN_{\rm ch}/\rmf d\eta$ with centrality cut 0-6\% at $\sqrt{s_{\rm NN}} = 200$ GeV \cite{Back:2001ae}. They are compared with $\rmf dn/\rmf d\eta$ at $\sqrt{s_{\rm NN}} = 130$ GeV in Fig. \ref{fig5}. Roughly speaking, the scaling on $\rmf dn/\rmf d\eta$ holds between $\sqrt{s_{\rm NN}} = 130$ GeV and 200 GeV\footnote{
Using the Bjorken's picture \cite{Bjorken:1983qr} for the calculation of energy density near $|\Delta \eta|\leq 0.5$ with the geometrical picture of the gold ($R_{\tau} \approx$ 6-7 fm, $c\tau_0 \approx$ 1-2 fm, $V \approx \pi R_{\rm T}^2(c\tau_0) \approx 300\ {\rm fm^3}$), we obtain the following values
\begin{eqnarray*}
  \varepsilon &\sim& \left. \frac 32 \frac 1V \frac{\rmf dN_{\rm ch}}{\rmf d\eta}E_{\rm T} \right|_{|\Delta\eta|\leq 0.5} \sim 1\ {\rm GeV/fm^3}\quad (130\ {\rm GeV})\, ,\\
  \varepsilon &\sim& 1.2\ {\rm GeV/fm^3}\quad (200\ {\rm GeV})\, .
\end{eqnarray*}
}. 
The distribution of $\rmf dN_{\rm ch}/\rmf d\eta$ or $\rmf dn/\rmf d\eta$ with the centrality cut 0-6\% does not show the particular behavior relating to the QGP in the sense of average. To investigate the particular phenomena like the turbulence and/or deflagration in $\rmf dN_{\rm ch}/\rmf d\eta$, we need to analyse the single event with smaller centrality cut than 0-6\%.

Moreover, analyses of event-by-event, for example the intermittency and the wavelet, seem to be necessary for $\rmf dN_{\rm ch}/\rmf d\eta$ with smaller centrality cut \cite{Burnett:1983pb,Takagi:1984gr,Bialas:1986jb,Biyajima:1990kw,Andreev:1995rc,Hwa:1990xg,Suzuki:1995kg}.

%
\section*{Acknowledgements}
One of authors (M. B.) is partially indebted to the Japanese Grant-in-aid for Education, Science, Sports and Culture (No. 09440103).

%

%
\end{document}